\documentclass[pre,aps,amsmath,longbibliography,amsfonts,amssymb,twocolumn]{revtex4-1}
\usepackage{bm}
\usepackage{graphicx,CJK}
\usepackage{dcolumn}
\DeclareMathOperator{\sgn}{sgn}
\begin{document}
\begin{CJK*}{UTF8}{mj}
\title{Branching annihilating random walk with long-range repulsion: logarithmic scaling, reentrant phase transitions, and crossover behaviors}
\author{Su-Chan Park (박수찬)}
\affiliation{Department of Physics, The Catholic University of Korea, Bucheon 14662, Republic of Korea}
\begin{abstract}
We study absorbing phase transitions in the one-dimensional branching annihilating random walk with 
long-range repulsion. The repulsion is implemented as hopping bias 
in such a way that a particle is more likely to hop away from its closest particle.
The bias strength due to long-range interaction has the form $\varepsilon x^{-\sigma}$, where $x$ is the distance from a particle to its closest particle, $0\le \sigma \le 1$, and the sign of $\varepsilon$ determines
whether the interaction is repulsive (positive $\varepsilon$) or attractive (negative $\varepsilon$). 
A state without particles is the absorbing state.
We find a threshold $\varepsilon_s$ such that the absorbing state is dynamically stable for small branching rate $q$ if $\varepsilon < \varepsilon_s$. 
The threshold differs significantly, depending on parity of the number $\ell$ of offspring.
When $\varepsilon>\varepsilon_s$, 
the system with odd $\ell$ can exhibit reentrant phase transitions from the active phase with nonzero steady-state density to the absorbing phase, and back to the active phase.
On the other hand, the system with even $\ell$ is in the active phase for nonzero $q$ if $\varepsilon>\varepsilon_s$.
Still, there are reentrant phase transitions for $\ell=2$.
Unlike the case of odd $\ell$, however, the reentrant phase transitions can occur only for $\sigma=1$ and $0<\varepsilon < \varepsilon_s$.
We also study the crossover behavior for $\ell = 2$ when the interaction is attractive (negative $\varepsilon$), to find the crossover exponent $\phi=1.123(13)$ for $\sigma=0$.
\end{abstract}
\date{\today}
\maketitle
\end{CJK*}
\section{\label{Sec:intro}Introduction}
The branching annihilating random walk (BAW)~\cite{Bramson1985,Sudbury1990} 
is a nonequilibrium reaction-diffusion system with pair annihilation, 
$A + A\rightarrow \emptyset$, and branching, $A \rightarrow (\ell + 1) A$.
Throughout this paper, we reserve $\ell$ to denote the number of offspring.
In a typical setting, particles diffuse symmetrically on a $d$-dimensional lattice and
pair annihilation occurs when two particles happen to occupy the same site.
The vacuum state without particles is an absorbing state in that
probability current from other states to the vacuum state is nonzero, while transition from
the vacuum state to any other state is prohibited.

If branching rarely occurs, then the steady-state particle density (of an infinite system)
is zero, which is a key characteristic of an absorbing phase.
For sufficiently large branching rate, 
the system can be in an active phase with nonzero steady-state particle density.
The absorbing phase transition from the absorbing phase to the active phase 
occurs at a certain critical branching rate.
In this paper, we are interested in critical behaviors of one-dimensional systems 
and in the following the dimension is always assumed 1.

The universality class to which the BAW belongs differs
according to parity of $\ell$~\cite{TT1992,J1993bawo,Jensen1993JPA,J1994}.
The BAW with odd $\ell$ belongs to the directed percolation (DP) universality class, 
as is consistent with the DP conjecture~\cite{J1981,G1982}.
The BAW with even $\ell$ can be mapped to a kinetic Ising model with two symmetric absorbing states~\cite{M1994} (see Sec.~\ref{Sec:nbawa} for a mapping) and belongs to the directed Ising (DI) universality class~\cite{HKPP1998}.
For a review of these two universality classes, see, e.g., Ref.~\cite{H2000,O2004,HHL2008Book}.
In the literature, the DI class is also called  
the parity-conserving universality class~\cite{GKvdT1984,J1993,CT1998},
DP2 class~\cite{H2000}, or the generalized voter class~\cite{HCDM2005}, depending on
which property of this universality class is emphasized. 

Recently, a modified version of the BAW was suggested by introducing hopping bias in such a
way that a particle is attracted by its closest particle~\cite{DR2019,Park2020B}.
The strength of the bias depends on the distance $x$ between a particle and its closest particle
in a power-law fashion $x^{-\sigma}$.
For odd $\ell$, this bias does not alter the critical behavior~\cite{DR2019}.
On the other hand, for even $\ell$, the modified model in one dimension does not belong to the DI class 
if $\sigma <1$. Rather, the critical exponents vary with $\sigma$~\cite{Park2020B}.
In Ref.~\cite{Park2020A}, the different critical behavior for $\sigma<1$ from the DI class
is attributed to the long-range nature of the bias, by studying the crossover that occurs
for large but finite range of attraction.
In this paper, we further generalize the BAW by allowing the long-range interaction to be repulsive
and study its critical phenomena.

The structure of this paper is as follows.
The definition of the generalized model will be provided in Sec.~\ref{Sec:model}.
If particles repel each other, then 
pair annihilation becomes inefficient to remove particles~\cite{Pun}.
Accordingly, one may anticipate that the model with repulsion is always in the active phase
for nonzero branching rate.
This anticipation turns out to be right if $\ell$ is even.
However, a careful analysis to be presented in Sec.~\ref{Sec:smallp} shows that 
the critical branching rate can be nonzero for odd $\ell$.
Rather unexpectedly, the model exhibits intriguing reentrant phase transitions, 
which will be analyzed in Sec.~\ref{Sec:odd} for $\ell=1$ and in Sec.~\ref{Sec:even} for $\ell=2$. 
The results are summarized in Sec.~\ref{Sec:sum}.

\section{\label{Sec:model}Model}
This section defines the model to be studied and summarizes established results 
that are relevant to later discussion.
We consider a one-dimensional lattice of size $L$ with periodic boundary conditions.
Each site is either occupied by at most one particle or vacant. 
A particle branches with rate $q$ or hops to one of its nearest neighbors with rate $1-q$
($0\le q \le 1$).  When a particle at site $i$ branches, its $\ell$ offspring are placed at 
consecutive sites $i+\xi, i+2\xi, \ldots, i+ \ell \xi$,
where $\xi$ is a random variable that takes $1$ or $-1$ with equal probability.
When a particle at site $i$ hops, it moves to site $i\pm 1$ with probability $(1\pm B_i)/2$, where 
\begin{align}
\label{Eq:Wdef}
	B_i &= \varepsilon\sgn(R_i-L_i) \left (\min\{R_i,L_i\}+\mu\right )^{-\sigma}, \\
\nonumber
        R_i &\equiv \min\{x|s_{i+x}=1,\, 1\le x \le L\},\\
\nonumber
        L_i &\equiv \min\{x|s_{i-x}=1,\, 1\le x \le L\}.
\end{align}
In the above equation, $\sgn(x)$ is the sign of $x$ with $\sgn(0)=0$, $\sigma$ is nonnegative, and 
$\varepsilon$, $\mu$ are constants with restriction $ |\varepsilon| < (1+\mu)^\sigma$ to ensure
$-1<B_i<1$.
If two particles happen to occupy a same site by any transition event,
then pair annihilation ($A+A\rightarrow \emptyset$) of the two particles occurs immediately.

The stochastic dynamics of the model can be represented as
\begin{align}
	\nonumber
	&1_i \prod_{k=1}^\ell c_{i+k} \rightarrow 1_{i} \prod_{k=1}^\ell \bar c_{i+k} \text{ with rate } \frac{q}{2},\\
	\nonumber
	&1_i \prod_{k=1}^\ell c_{i-k} \rightarrow 1_{i} \prod_{k=1}^\ell \bar c_{i-k} \text{ with rate } \frac{q}{2},\\
	\nonumber
	&1_i c_{i+1} \rightarrow 0_i \bar {c}_{i+1} \text{ with rate } (1-q) \frac{1+B_i}{2},\\
	&1_i c_{i-1} \rightarrow 0_i \bar {c}_{i-1} \text{ with rate } (1-q) \frac{1-B_i}{2},
	\label{Eq:rule}
\end{align}
where $c_j$ stands for the occupation number at site $j$ with 
$\bar{c}_j \equiv 1-c_j$ and $1_i$ ($0_i$) means that
site $i$ is occupied (vacant).
Monte Carlo simulations for the rule~\eqref{Eq:rule} were performed in the following way.
Assume that there are $M$ particles at time $t$. 
Choose one among $M$ particles at random with equal probability.
Assume that a particle at site $i$ is chosen.
After  the choice, we generate a random number $\theta$ that is uniformly distributed
in $0<\theta<1$.
If $\theta < q/2$, then $\ell$ offspring are placed on the right-hand side
of $i$.
Else if $\theta < q$, then $\ell$ offspring are placed on the left-hand side
of $i$.
Else if $\theta < q + (1-q)(1+B_i)/2$, then
the chosen particle hops to the right.
Otherwise, the chosen particle hops to the left.
Pair annihilation occurs if applicable. 
After the change of a configuration, time increases by $1/M$.

The model will be called the BAW with long-range interaction (BAWL).
For positive (negative) $\varepsilon$,
a particle is in a sense repulsed (attracted) by its closest particle located at $\min\{R_i,L_i\}$.
Thus, we will refer to the model with positive (negative) $\varepsilon$ as the BAW with long-range \emph{repulsion} (\emph{attraction})
to be abbreviated as BAWLR (BAWLA).

We define the particle density $\rho(t)$ as
\begin{align}
	\rho(t)  = \lim_{L\rightarrow \infty} \frac{1}{L} \sum_{i=1}^L \langle c_i \rangle,
\end{align}
where $c_i$ is the occupation number at site $i$ at time $t$ and
$\langle \cdots \rangle$ stands for average over ensemble.
In actual simulations, the system size $L$ is so large that a finite-size effect is negligible 
up to the observation time and the system evolves from the 
fully-occupied initial condition with $\rho(0)=1$.

%
%
The BAWL with $q=0$ is the annihilating random walk with long-range interaction (AWL), extensively studied in Ref.~\cite{Pun} (see also Refs.~\cite{Sen2015,Park2020B} for the AWL with 
attractive interaction).
We summarize some results of Ref.~\cite{Pun} that are relevant to later discussion.
For the two-particle initial condition 
in which there are only two particles in a row at $t=0$, 
the probability $P_s$ that the two particles are never annihilated is 
\begin{align}
	P_s^{-1} = 1 + \sum_{i=1}^{\infty} \prod_{k=1}^i \frac{1-\varepsilon(k+\mu)^{-\sigma}}{1+\varepsilon (k+\mu)^{-\sigma}}.
	\label{Eq:Ps}
\end{align}
For $\sigma=0$ and $\sigma=1$, $P_s$ becomes
\begin{align}
	\label{Eq:Pss0}
	P_s(\sigma=0) &= \frac{2\varepsilon}{1+\varepsilon}\Theta(\varepsilon),\\
	\label{Eq:Pss1}
	P_s(\sigma=1) &= \frac{2\varepsilon-1}{\mu+\varepsilon}\Theta(2 \varepsilon-1),
\end{align}
where  $\Theta(\cdot)$ is the Heaviside step function.
For $0<\sigma<1$, $P_s$ is nonzero for $\varepsilon>0$ and 0 for $\varepsilon\le 0$.
When necessary, one can numerically calculate $P_s$ for $0<\sigma<1$ using Eq.~\eqref{Eq:Ps}.

When there are only three particles at sites $i-r$, $i$, $i+r$ and particles at $i\pm r $ are assumed
immobile, the time $T_0(r)$ taken for the particle in the middle to encounter one of the two particles at
$i \pm r$ behaves for large $r$ as
\begin{align}
	T_0 &\sim
	\begin{cases}
		r^\sigma \exp \left ( C_\sigma r^{1-\sigma} \right ), & \sigma<1,\\
		r^{1+2\varepsilon}, & \sigma=1 \quad \&\quad  2\varepsilon > 1,\\
		r^2 \ln r , & \sigma=1\quad  \& \quad 2\varepsilon = 1,\\
		r^2, & \text{ otherwise.}
	\end{cases}
	\label{Eq:T0posi}\\
	C_\sigma &\equiv \begin{cases}
                \ln[(1+\varepsilon)/(1-\varepsilon)], &\sigma=0,\\
                2\varepsilon/(1-\sigma), & 0<\sigma<1,\\
        \end{cases}
\end{align}
for positive $\varepsilon$ and
\begin{align}
	T_0 \sim \begin{cases}
		r^{1+\sigma}, & \sigma\le 1,\\
		r^2, &\sigma > 1,
	\end{cases}
\end{align}
for negative $\varepsilon$.

Interpreting $2/T_0$ with $r=1/\rho(t)$ as the rate of the density decrease per particle at
time $t$, it was found~\cite{Pun} that 
\begin{align}
	\rho(t) \sim
	\begin{cases} (\ln t)^{-1/(1-\sigma)}, & \sigma<1\quad\& \quad \varepsilon >0,\\
		t^{-1/(1+2\varepsilon)}, & \sigma =1\quad\& \quad 2\varepsilon>1,\\
		t^{-1/2}, &\text{ otherwise,}
	\end{cases}
	\label{Eq:rho0}
\end{align}
for $\varepsilon>0$ and 
\begin{align}
	\rho(t) \sim
	\begin{cases} t^{-1/(1+\sigma)}, & \sigma \le 1,\\
		t^{-1/2}, & \sigma \ge 1,
	\end{cases}
\end{align}
for $\varepsilon<0$.

We would like to comment on a recent numerical work~\cite{Roy2020}
of the AWL for $\sigma=0$ and positive $\varepsilon$,
which claims that  $\rho(t)$ behaves as $(\ln t)^{-b}$ with
$b$ to vary continuously, depending on $\varepsilon$. 
We think, however, the neglect of $\ln \ln t$ correction~\cite{Pun} errorneously gives this conclusion.


\section{\label{Sec:smallp} Stability of the absorbing state}
Since $P_s$ for $\sigma<1$ and $\varepsilon>0$ is positive as seen in Sec.~\ref{Sec:model},
an offspring has a chance to avoid being annihilated with its parent, which might lead us
to claim that the critical point of the BAWLR with $\sigma<1$ is zero 
and the system is in the active phase as long as $q$ is nonzero.
To check the validity of this claim, we investigate the stability of the absorbing state 
for small $q$. 

To this end, we consider an initial condition that each site is occupied with
probability $\rho_0$. Assuming both $\rho_0$ and $q$ are small ($0< \rho_0 \ll 1, q\ll 1$),
we can treat any particle-number changing event (branching or pair annihilation) as
independent, because a typical distance between sites at which two consecutive such events in time domain
occur should be large in space.
We first consider the case with $\ell=1$ and general cases will be discussed later.

The rate of density decrease due to pair annihilation is roughly $2/T_0(r)$
with $r=1/\rho_0$~\cite{Pun}.
To estimate the rate of density increase by branching, we assume that a particle at site
$i$, to be called the parent, begets its offspring at $t=0$.  With probability $P_s$ the offspring
will escape from the parent and the number of particles increases by 1, while
with probability $1-P_s$ the parent and the offspring undergo pair annihilation, which makes 
the number of particles decrease by 1.  Hence, the rate of density increase by branching is
$q [P_s - (1-P_s)] = q(2 P_s-1)$. 
The independence argument finally gives
\begin{align}
	\frac{d\ln \rho}{dt} \approx q\left (2P_s-1\right ) - \frac{2}{T_0(1/\rho)}.
	\label{Eq:drdt}
\end{align}

If $P_s<1/2$, then the branching triggers even faster (exponential) density decay,
which makes the absorbing state stable.
On the other hand, if $P_s>1/2$, then the absorbing state is unstable, because 
$1/T_0 \rightarrow 0$ as $\rho_0\rightarrow 0$; see Eq.~\eqref{Eq:T0posi}.
Hence the condition for the absorbing state to be stable is $P_s<1/2$. 
As we will see soon, the condition will differ significantly for $\ell=2$.

Unlike the claim at the beginning of this section, there should be a positive threshold $\varepsilon_s$ 
determined by $P_s(\varepsilon_s) = \frac{1}{2}$ such that for $0<\varepsilon < \varepsilon_s$,
the steady-state density for small $q$ is zero
and, in turn, the critical point $q_c$ should be nonzero.
From Eq.~\eqref{Eq:Ps}, $\varepsilon_s$ is found as
	\begin{align}
		\varepsilon_s = 
		\begin{cases} 1/3, & \sigma=0,\\
			0.438~213~752, & \sigma=0.5,~\mu=1,\\
			(\mu+2)/3, & \sigma=1,
		\end{cases}
	\label{Eq:eth}
	\end{align}
where the numerical value of $\varepsilon_s$ for $\sigma=0.5$ is presented for later purpose.

When $P_s<1/2$, Eq.~\eqref{Eq:drdt} suggests that the density decreases
exponentially for small but nonzero $q$, which is a typical feature
of the absorbing phase of systems belonging to the DP class.
This is usually interpreted as the generation of spontaneous annihilation 
by the chain of reaction $A\rightarrow 2A \rightarrow 0$.
Hence, we anticipate that the BAWL with $\ell=1$ belongs to the DP class if $P_s < 1/2$.

\begin{figure}
        \includegraphics[width=\linewidth]{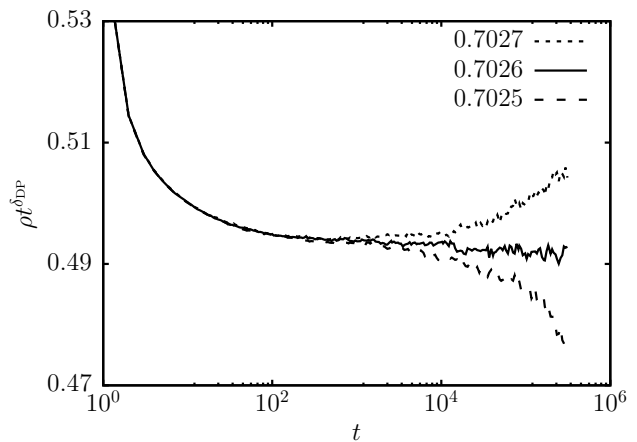}
        \caption{\label{Fig:dp} Semilogarithmic plots of
        $\rho t^{-\delta_\textrm{DP}}$ vs. $t$ for
        $\sigma=0.8$, $\varepsilon=0.5$, $\mu=0$, and $\ell=1$ with $q=0.7025$, 0.7026, and 0.7027 (bottom to top),
        where $\delta_\textrm{DP} \approx 0.159~46$ is the critical decay
        exponent of the DP class.
        The curve for $q=0.7026$ is flat for more than two logarithmic decades,
        while the other curves eventually veer up (0.7027) or down (0.7025), indicating
        that the transition point is $q_c = 0.7026(1)$
        and this case belongs to the DP class.
        }
\end{figure}
To confirm this scenario, we performed simulations for $\sigma=0.8$, $\varepsilon = 0.5$, and $\mu=0$
with $L= 2^{22}$.
$P_s$ for this case is about 0.445. 
If this case does indeed belong to the DP class,
then the density at the critical point should decay as
\begin{align}
	\rho(t) \sim t^{-\delta_\text{DP}},
\end{align}
with the critical decay exponent $\delta_\text{DP} = 0.159~46$~\cite{J1999} of the DP class.
In Fig.~\ref{Fig:dp}, we depict
$\rho t^{\delta_\textrm{DP}}$ versus $t$ on a semilogarithmic scale
for $q=0.7025, 0.7026$, and 0.7027. 
To reduce statistical error, we averaged over 40 independent runs for each parameter set.
The curve for $q=0.7025$ (0.7027) eventually veers down (up), while
the curve for $q=0.7026$ remains flat for long time. Therefore, we conclude that 
the critical point is $q_c = 0.7026(1)$, where the number in parentheses indicates
the uncertainty of the last digit, and this case belongs to the DP class as anticipated.

Since the absorbing state is unstable if $P_s > 1/2$, 
the steady state should have nonzero density for small but nonzero $q$.  
Setting the time derivative of $\rho$ in Eq.~\eqref{Eq:drdt} to be 0, 
the steady-state density $\rho_s$ can be estimated by
\begin{align}
	T_0(1/\rho_s) \sim 1/q,
	\label{Eq:ss}
\end{align}
which, along with Eq.~\eqref{Eq:T0posi}, gives
\begin{align}
	\rho_s \sim
	\begin{cases}
		(-\ln q)^{-1/(1-\sigma)}, & \sigma<1,\\
		q^{1/(1+2\varepsilon)}, & \sigma=1,\\
	\end{cases}
\end{align}
for small $q$ and $\varepsilon > \varepsilon_s$. Hence, there is a continuous transition with $q_c = 0$.

\begin{figure}
	\includegraphics[width=\linewidth]{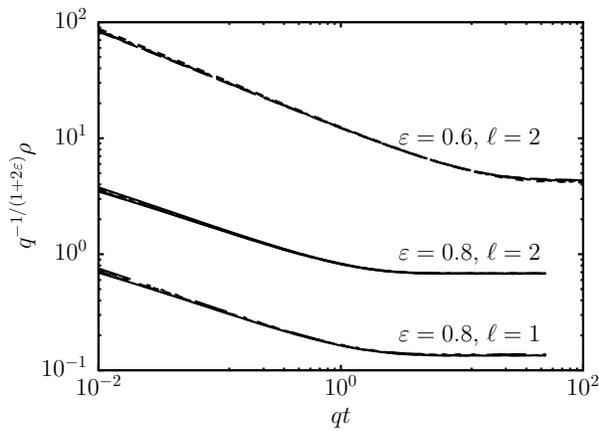}
	\caption{\label{Fig:s1col} Data collapse plots of
	$q^{-1/(1+2\varepsilon)} \rho$ vs. $qt$ for
	$\varepsilon=0.6$, $\ell=2$ (top), $\varepsilon=0.8$, $\ell=2$ (middle),
	and $\varepsilon=0.8$, $\ell=1$ (bottom) on a double logarithmic scale.
	Here, $\sigma$ is set 1.
	$q$ for $\varepsilon=0.6$ ranges from $10^{-5}$ to $5\times 10^{-5}$
	and $q$ for $\varepsilon = 0.8$ (for both $\ell=1$ and $\ell=2$)
	ranges from $10^{-4}$ to $5\times 10^{-4}$.
	To distinguish collapse plots from each other,
	we multiply $\rho$ by an arbitrary constant for each parameters set.
	}
\end{figure}
Now, we make an ansatz for a scaling function that describes the absorbing phase transition with $q_c=0$.
From the above discussion, $1/q$ should be a characteristic time scale of the system.
For $\sigma=1$, the density would be described by a scaling function of the form
\begin{align}
	\rho(t;q) = q^{1/(1+2\varepsilon)} G_1(q t).
	\label{Eq:s1scale}
\end{align}
To ensure that $\rho(t;q)$ has the desired behavior as $t\rightarrow \infty$, 
the scaling function $G_1$ should have the following asymptotic behavior:
$G_1(x) \sim 1$ as $x\rightarrow \infty$ and $G_1(x) \sim x^{-1/(1+2\varepsilon)}$ as
$x \rightarrow 0$ (actually, the limit $x\rightarrow 0$ should be understood as a joint limit
$t\rightarrow \infty$ and $qt\rightarrow 0$).

For $\sigma < 1$, the logarithmic behavior should be incorporated properly into the scaling function.
Since the characteristic time scale is $1/q$ and there is a logarithmic density decay at $q=0$, 
a proper scaling parameter should be $-\ln t/\ln q$. Thus, we write
\begin{align}
	\rho(t;q) = \left ( - \ln q \right )^{-1/(1-\sigma)} G_\sigma(-\ln t/\ln q),
	\label{Eq:scale}
\end{align}
for $\sigma<1$.
The desired asymptotic behaviors for large $t$ are reproduced if 
$G_\sigma(x) \sim x^{-1/(1-\sigma)}$ for small $x$ and $G_\sigma(x) \sim 1$ for large $x$.
Notice that if we choose $qt$ as a scaling parameter as in Eq.~\eqref{Eq:s1scale},
we cannot reproduce the desired asymptotic behavior in the small $qt$ limit.

Before confirming the validity of the scaling ansatz by numerical simulations,
we will discuss the stability of the absorbing state for $\ell>1$.

Right after a particle branches, it and its offspring form a cluster of size $\ell + 1$. 
Particles in the middle of the cluster are likely to be annihilated in short time.
If $\ell$ is odd, only two particles in the cluster are likely to remain in short time 
and the fate of the cluster is almost identical to the case of $\ell=1$.
Hence the scenario for $\ell=1$ is applicable to the BAWL with odd $\ell$, though 
the value of $\varepsilon_s$ can differ depending on $\ell$.

If $\ell$ is even, then the size of the cluster shrinks to 3 in short time.
By $\tilde P_s$, we denote the probability that no pair annihilation happens in the AWL 
if the system evolves from a configuration only with three particles in a row.
That is, with probability $\tilde P_s$ all the three particles survive and
with probability $1-\tilde P_s$ only one particle survives.
Since at least one particle always survives, the rate of the density increase by branching is 
$2 \tilde P_s$, which gives
\begin{align}
	\frac{d\ln \rho}{dt} \approx 2 q \tilde P_s - \frac{2}{T_0}.
	\label{Eq:smallqDI}
\end{align}

\begin{figure}
	\includegraphics[width=\linewidth]{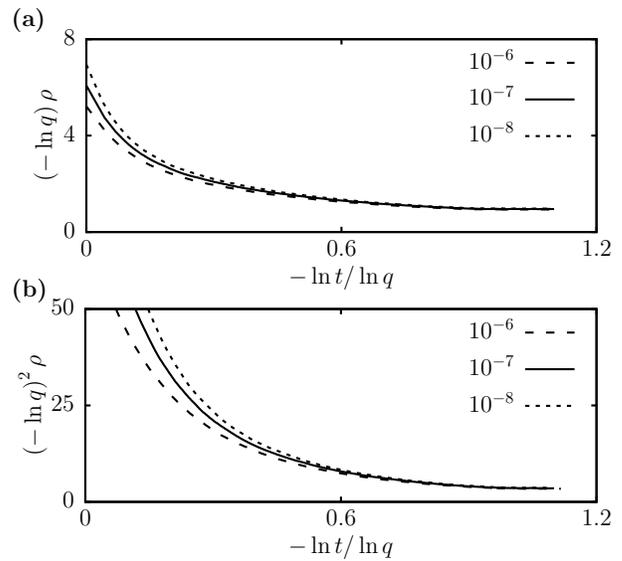}
	\caption{\label{Fig:log} Plots of
	$\left ( -\ln q \right )^{1/(1-\sigma)} \rho$ vs
	$-\ln t/\ln q$ for (a) $\sigma=0$, $\ell=1$ and (b) $\sigma=0.5$, $\ell=2$.
	Here, we set $\mu=0$ and $\varepsilon=0.5$.
	For large $t$, all data are well collapsed onto a single curve.
	}
\end{figure}
Now we discuss the relation between $P_s$ and $\tilde P_s$.
Since the particle in the middle is more likely to be annihilated than 
particles in the two-particle initial condition, $\tilde P_s$ cannot be larger than $P_s$.
In the mean time, if $P_s > 0$, the three particles should survive with nonzero probability once
they happen to be separated by a large distance. Since probability of being separated by a large distance
cannot be zero as far as the distance is finite, $P_s>0$ should entail $\tilde P_s > 0$.
Therefore, $\tilde P_s$ is positive if and only if $P_s$ is positive. 
Accordingly, if $P_s$ is positive and $\ell$ is even, then the absorbing state is unstable and
the scaling functions \eqref{Eq:s1scale} and \eqref{Eq:scale} are expected to describe
the critical phenomena.
If $P_s=\tilde P_s=0$ and $q$ is small, then 
Eq.~\eqref{Eq:smallqDI} shows that
only pair annihilation governs the long-time behavior of the particle density.

To confirm the scaling ansatz as well as the scenario for even $\ell$, 
we performed Monte Carlo simulations for two cases, $\ell=1$ and $\ell=2$.
First, we present simulations results for $\sigma=1$ and $\mu=0$.
Figure~\ref{Fig:s1col} depicts data-collapse plots using Eq.~\eqref{Eq:s1scale}
for $\varepsilon=0.6$ ($\ell=2$) and $\varepsilon=0.8$ ($\ell=1$ and $\ell=2$).
Notice that $\varepsilon_s$ for $\ell=1$ and $\mu=0$ is $\frac{2}{3}\approx 0.67$.
The system size in the simulations is $L=2^{20}$ and number of independent runs ranges from 8 to 40.
The collapse is almost perfect, which confirms 
the scaling ansatz~\eqref{Eq:s1scale}.

Next, we present simulation results for $\sigma=0$ ($\ell=1$) and $\sigma=0.5$ ($\ell=2$). 
For both cases, we set $\mu=0$ and $\varepsilon =0.5$.
The values of $q$ are $10^{-8}$, $10^{-7}$, and $10^{-6}$. 
The system sizes in the simulations are $2^{14}$ and $2^{15}$ for $\sigma=0$ and $\sigma=0.5$, respectively,
The number of independent runs is 8 for all cases.
Figure~\ref{Fig:log} depicts data-collapse plots according to the scaling ansatz~\eqref{Eq:scale},
which shows a nice agreement.

\begin{figure}
	\includegraphics[width=\linewidth]{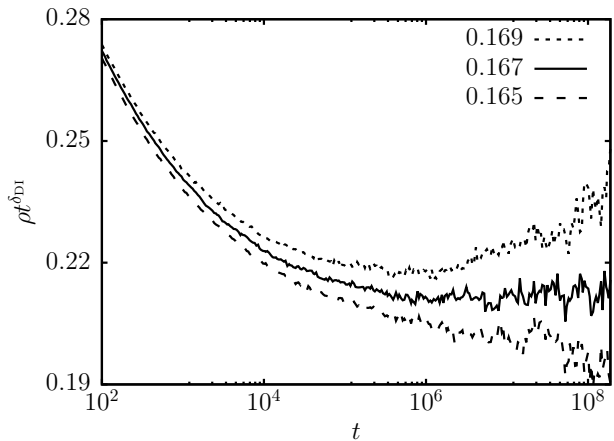}
	\caption{\label{Fig:dis1} Semilogarithmic plots of
	$\rho t^{\delta_\text{DI}}$ vs.
	$t$ for $q=0.165$, $0.167$, and $0.169$, bottom to top.
	The other parameters are $\ell=2$, $\sigma=1$, $\mu=0.2$, and $\varepsilon = 0.35$.
	$\delta_\text{DI}$ is the critical decay exponent of the DI class with numerical
	value $0.2872$. The curve for $q=0.167$ is flat for more than two logarithmic decades,
        while the other curves eventually veer up (0.169: top curve) or down (0.165 : bottom curve), 
	indicating that the transition point is $q_c = 0.167(2)$
        and this case belongs to the DI class.
	}
\end{figure}
Now we discuss absorbing phase transitions of the BAWLR with $\ell = 2$ for $\sigma=1$ and
$0<\varepsilon < 0.5$. In this regime, $P_s = 0$ and
the density decays as $t^{-1/2}$ for small $q$; see Eq.~\eqref{Eq:smallqDI} along with Eq.~\eqref{Eq:rho0}.
Hence, it can be anticipated that there should be a phase transition at nonzero $q_c$ and
this case should belong to the DI class.

To confirm, we simulated the system of size $L=2^{21}$
with $\sigma=1$, $\mu=0.2$, and $\varepsilon = 0.35$ for
$q=0.165$, $0.167$, and $0.169$. For each parameter set, we performed
80 independent runs for average.
In Fig.~\ref{Fig:dis1}, we depict $\rho t^{\delta_\text{DI}}$ as a function of $t$ on a
semilogarithmic scale, where $\delta_\text{DI} \approx 0.2872$~\cite{Park2013}
is the critical decay exponent of the DI class.
Since the graph for $q=0.165$ (0.169) eventually veers down (up),
we conclude that the critical point is $q_c = 0.167(2)$ and, as expected, this case belongs to the DI class.

\section{\label{Sec:odd} Single offspring}
In this section, we continue studying absorbing phase transitions of the BAWLR with $\ell=1$.
When it comes to the universal behavior,
nothing new beyond the result in Sec.~\ref{Sec:smallp} is expected.
Rather, we are interested in the behavior of the phase boundary
as $\varepsilon$ approaches $\varepsilon_s$.

To this end, we performed Monte Carlo simulations for $\sigma\le 1$.
Differently from the numerical studies in Sec.~\ref{Sec:smallp}, 
$q$ is now fixed and $\varepsilon$ is a tuning parameter.
The critical point will be denoted by $\varepsilon_c$. 
Since the model belongs to the DP class, 
the critical point was found using the method as in Fig.~\ref{Fig:dp} (details not shown here).

\begin{table}
	\caption{\label{Tab:ec} Critical points $\varepsilon_c$ of the BAWL with $\ell=1$ 
	for $\sigma=0$ ($\mu=0$; second column), $0.5$ ($\mu=1$; third column), and $1$ ($\mu=1$; fourth column). Numbers in parentheses indicate uncertainty of the last digits.}
\newcolumntype{.}{D{.}{.}{8}}
\newcolumntype{,}{D{.}{.}{4}}
\begin{ruledtabular}
	\begin{tabular}{,...}
		q&\multicolumn{3}{c}{$\varepsilon_c$}\\
		\cline{2-4}
		&\multicolumn{1}{c}{$\sigma=0$}&\multicolumn{1}{c}{$\sigma=0.5$}&\multicolumn{1}{c}{$\sigma=1$}\\
		\hline
0.0001&    0.3335(1)&    0.4447(1)&    1.0845(5)\\
0.0003&    0.3339(1)&  0.452~85(5)&    1.1035(2)\\
 0.001&    0.3356(1)&  0.469~75(2)&    1.1285(1)\\
 0.003&    0.3419(1)&    0.4933(1)&    1.1553(1)\\
  0.01&  0.363~40(2)&  0.526~24(3)&    1.1883(1)\\
  0.03&  0.403~08(3)&  0.559~24(2)&    1.2199(1)\\
   0.1&    0.4639(1)&    0.5923(1)&    1.2498(1)\\
   0.2&  0.500~05(4)&    0.6044(1)&    1.2559(1)\\
   0.3&  0.516~35(5)&  0.604~43(3)&  1.246~15(5)\\
   0.4&    0.5207(1)&  0.596~05(5)&    1.2228(1)\\
   0.5&  0.513~65(5)&    0.5780(1)&    1.1814(1)\\
   0.6&    0.4911(1)&    0.5450(1)&    1.1109(1)\\
   0.7&    0.4410(1)&    0.4838(1)&    0.9838(1)\\
   0.8&    0.3251(1)&    0.3531(1)&  0.716~65(5)\\
   0.9&   -0.0496(1)&   -0.0533(1)&   -0.1081(1)
	\end{tabular}
\end{ruledtabular}
\end{table}
We simulated the BAWLR for three sets of parameters, $\{\sigma=0, \mu=0\}$, $\{\sigma=0.5,
\mu = 1\}$, and $\{\sigma=1, \mu=1\}$.
The threshold values for these parameter sets are given in Eq.~\eqref{Eq:eth}.
The critical points are tabulated in Table~\ref{Tab:ec}. 
It is intriguing that $\varepsilon_c$ approaches $\varepsilon_s$ from above. 
Since $\varepsilon_c$ becomes smaller than $\varepsilon_s$ for sufficiently large $q$,
there should be a reentrant phase transition
as $q$ varies with $\varepsilon$ to be fixed slightly larger than $\varepsilon_s$.

\begin{figure}
	\includegraphics[width=\linewidth]{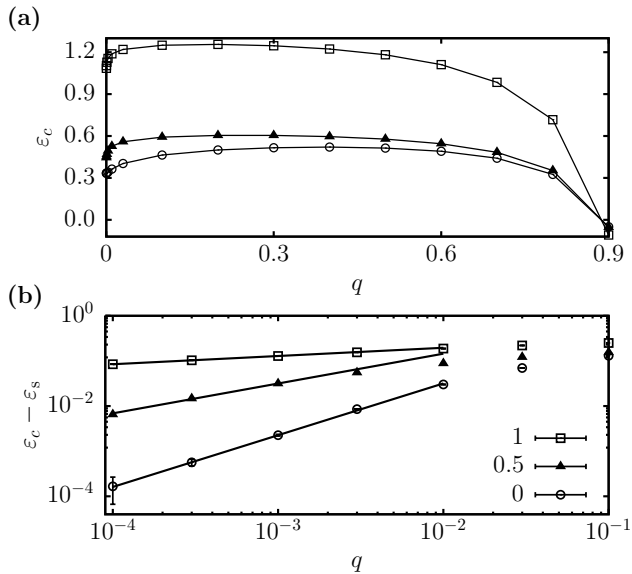}
	\caption{\label{Fig:dpb} (a) Plots of $\varepsilon_c$ vs. $q$ for $\sigma=0$ (circle),
	$\sigma=0.5$ (triangle), and $\sigma=1$ (square), bottom to top. 
	Error bars are hardly observable in this scale.
	(b) Log-log plots of $\varepsilon_c-\varepsilon_s$ vs. $q$ for $\sigma=0$ (circle),
        $\sigma=0.5$ (triangle), and $\sigma=1$ (square), bottom to top. 
	Line segments with slope 0.18, 0.66, 1.15 (top to bottom) show
	the results of power-law fitting.
	}
\end{figure}

To illustrate the reentrant phase transitions,
we depict phase boundaries ($\varepsilon_c$ versus $q$) in Fig.~\ref{Fig:dpb}(a).
There are two salient features of the phase boundaries.
First, the reentrant phase transitions indeed occur for $\varepsilon_s < \varepsilon < \varepsilon_b$.
We roughly estimate $\varepsilon_b$ as
0.521, 0.605, and 1.26, for $\sigma=0$, $0.5$, and 1, respectively, from Table~\ref{Tab:ec}.
Second, the behavior of the phase boundary for small $q$ varies
with $\sigma$. To be concrete,
we analyzed how $\varepsilon_c$ approaches $\varepsilon_s$ as
$q\rightarrow 0$. In Fig.~\ref{Fig:dpb}(b),  $\varepsilon_c -\varepsilon_s$ is drawn against
$q$ on a double logarithmic scale, along with power-law fitting results for
small $q$. We find 
\begin{align}
	\varepsilon_c - \varepsilon_s \sim q^{\phi}
	\label{Eq:esphi}
\end{align}
with $\phi = 1.15$, $0.66$, and  $0.18$ for $\sigma=0$, 0.5, and 1, respectively.
This power-law behavior is reminiscent of a crossover behavior.

Now we provide a hand-waving argument as to why there are reentrant phase transitions.
Let us revisit the initial condition with low density as in Sec.~\ref{Sec:smallp}.
Assume that a particle at site $i$ branches its offspring at site $i+1$ at time $t=0$.
If these two particles survive, then next branching occurs approximately at time $1/q$.
We assume that the parent is at site $j$ and it branches a second
offspring at site $j+1$.
Since the mean distance of the surviving samples with the two-particle initial condition
increases as $t^{1/(1+\sigma)}$~\cite{Pun},
the average distance between the parent and its
first offspring at the time of the second birth is $d \approx q^{-1/(1+\sigma)}$. Now there are three particles at
site $j$, $j+1$, and $j+d$. 

Let $P_m(d)$ be the probability that three particles survive forever if 
the system (with $q=0$) evolves from the initial condition with only three particles at
sites $j$, $j+1$ and $j+d$.
Obviously, $P_m(d)$ should be an increasing function of $d$ with $P_m(d) \rightarrow P_s$ as 
$d\rightarrow\infty$. 
Accordingly, it is not impossible for $P_m(d)$ to be smaller than $\frac{1}{2}$ even if
$P_s$ is larger than $\frac{1}{2}$.

The rate $w$ of particle-number change due to the two branching events is
\begin{align}
	w&=-q(1-P_s) +  q P_s \left [ 1 -q (1 - P_m) + qP_m \right ]\nonumber \\
	&= q (2P_s-1) - q^2 P_s (1-2P_m),
\end{align}
which modifies Eq.~\eqref{Eq:drdt} as
\begin{align}
	\frac{d\ln \rho}{dt} \approx  q (2P_s-1) - q^2 P_s (1-2 P_m) - \frac{2}{T_0 }.
	\label{Eq:drdtn}
\end{align}
If $P_s>1/2$ and $q$ is extremely small but nonzero, the absorbing state is unstable 
as we have seen. Since $P_m$ is a decreasing function of $q$ (recall that $d$ is a decreasing function of $q$), 
$P_m$ may be smaller than $\frac{1}{2}$
from certain $q$ and the stability of the absorbing state can change at, say, $q_1$, which can be
 found approximately as a solution of 
\begin{align}
	q (1-2P_m)= 2P_s - 1,
	\label{Eq:con}
\end{align}
where $P_m$ should be regarded as a function of $q$.
Since the system should be in the active phase if $q$ is close to 1,  
the stability of the absorbing state changes again at, say, $q_2$ ($q_2>q_1$), which 
explains the existence of reentrant phase transitions.

In mathematics literature, a model is called \emph{attractive} if, roughly speaking, more particles induce 
longer life time (and/or larger chance of survival) of the system.
A typical example of an attractive model is the contact process~\cite{H1974,Liggett1985Book}.
The BAW is not attractive~\cite{Bramson1985} and neither is the BAWL.
Thus, larger branching rate does not necessarily imply more chance of survival.
In the BAWLR, we have shown that larger branching rate sometimes makes the system
fall into an absorbing state.
To our knowledge, the BAWLR is the first model that exhibits a reentrant phase transition
due to the lack of attractiveness. 
It is worth commenting that the reentrant phase transition observed in Ref.~\cite{KP1995}
should be attributed to the absence of the active phase in the branching annihilating process~\cite{Sudbury1990}, because the reentrant phase transition disappears when dynamic branching is employed~\cite{KP1995}.

\section{\label{Sec:even} two offspring}
In this section, $\ell$ is always set to be 2 and
we do not explicitly mention the value of $\ell$ when we refer to the model.
If $\sigma<1$ and $\varepsilon>0$ or if $\sigma=1$ and $2\varepsilon > 1$, that is, if $P_s>0$, then
the absorbing state is dynamically unstable and
the critical phenomena are completely described
by the scaling functions~\eqref{Eq:s1scale} and ~\eqref{Eq:scale}.
Hence, this section is interested in phase transitions in the cases with $P_s = 0$.

For $\sigma<1$, the condition $P_s = 0$ is applicable only to the BAWLA.
Recall that the BAWLA  with nonzero $\varepsilon$ does not belong to the DI class~\cite{DR2019,Park2020B}.
Thus, there should be a crossover behavior for small $|\varepsilon|$, which is the topic of
Sec.~\ref{Sec:nbawa}.

For $\sigma=1$, $P_s$ can be zero even if $\varepsilon >0$ and this case, as shown in Sec.~\ref{Sec:smallp},
belongs to the DI class. Quite unexpectedly, another reentrant phase transitions occur
in a certain region of $\varepsilon$.
These reentrant phase transitions will be studied in Sec.~\ref{Sec:nbawr2}.

\subsection{\label{Sec:nbawa}Crossover in the BAWLA with $\sigma=0$}
The crossover behavior is described by a scaling function
\begin{align}
	\rho(t;q,\varepsilon) = t^{-\delta_\text{DI}} R\left [t |\varepsilon|^{\nu_\|/\phi},  (q-q_0)t^{1/\nu_\|}\right],
	\label{Eq:cross}
\end{align}
where $q_0$ is the critical point for $\varepsilon=0$; $\delta_\text{DI}$ and $\nu_\|$ 
are the critical exponent of the DI class; and $\phi$ is the crossover exponent.
In a recent study~\cite{Park2020}, 
$\nu_\|$ is found to be 3.55 and we will use this value for a data-collapse plot.

\begin{figure}
	\includegraphics[width=\linewidth]{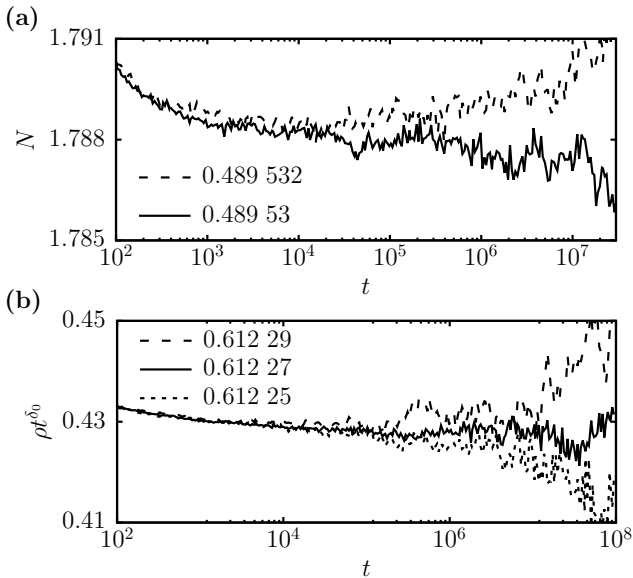}
	\caption{\label{Fig:di} (a) Semilogarithmic plots of $N$ vs. $t$ 
	for $\varepsilon=0$ and $\ell=2$ around the critical point. 
	The curve for $q=0.489~532$ (top)
	veers up for large $t$, while the curve for $q=0.489~53$ (bottom) veers down,
	which yields $q_0 = 0.489~531(1)$.
	(b) Semilogarithmic plot of $\rho t^{\delta_0}$ vs. 
	$t$ for the BAWLA with $\ell=2$ around the critical point 
	for $\sigma=0$ and $\varepsilon=-0.16$ with 
	$\delta_0 = 0.2394$. The curve for $q=0.612~29$ (top)
	veers up for large $t$, while the curve for $q=0.612~29$ (bottom) veers down.
	We conclude $q_c = 0.612~27(2)$.
	}
\end{figure}
To find the crossover exponent, we analyze how the phase boundary 
behaves when $|\varepsilon|$ is small.
If we denote the critical point for negative $\varepsilon$ by $q_c$,
Eq.~\eqref{Eq:cross} gives
\begin{align}
	|q_c-q_0| \sim |\varepsilon|^{1/\phi},
\end{align}
which will be used to find the crossover exponent $\phi$.

We first find $q_0$ for $\varepsilon = 0$.
In the literature, the value of $q_0$ is available 
(for example see Refs.~\cite{KP1995,PP2008PRE}),
but we will provide more accurate value in this paper. To compare our estimate to 
the literature, one should notice that 
$p$ is usually used to denote the hopping probability, which is $1-q$ in this paper.

To find $q_0$, we performed simulations with the two-particle initial condition.
At the critical point, the number $N(t)$ of particles averaged over all
ensemble behaves for large $t$ as
\begin{align}
	N(t) \sim t^\eta.
	\label{Eq:N}
\end{align}
Since $\eta\approx 0$ for the DI class~\cite{Park2013}, 
$N(t)$ drawn against $t$ should veer up (down)
if the system is in the active (absorbing) phase. 
Thus, we will find $q_0$ by plotting 
$N$ as a function of $t$ on a semilogarithmic scale around the critical point.

\begin{table}
	\caption{\label{Tab:pc} Critical points $q_c$ of the BAWLA 
	with $\ell=2$ for $\sigma=0$. Numbers in parentheses indicate uncertainty of 
	the last digits.}
\newcolumntype{.}{D{.}{.}{9}}
\begin{ruledtabular}
	\begin{tabular}{..}
		\varepsilon&q_c\\
		\hline
		0       &0.489~531(1)\\
		-0.0025  &0.4936(2)\\
		-0.005   &0.497~15(15)\\
		-0.01    &0.5036(1)\\
		-0.02    &0.5151(1)\\
		-0.04    &0.5346(1)\\
		-0.08    &0.5658(1)\\
		-0.16    &0.612~27(2)
	\end{tabular}
\end{ruledtabular}
\end{table}
In Fig.~\ref{Fig:di}(a), we present the simulation results. 
The number of independent runs for each parameter is $5\times 10^8$.
Since the graph for $q=0.489~532$ ($0.489~53$) veers up (down) in the long-time limit,
we conclude that $q_0 = 0.489~531(1)$. Note that this estimate is more accurate than 
previous estimates in the literature (for example, in Ref.~\cite{PP2008PRE}, the critical point
was estimated as $1-p_c = 0.489~65(5)$, but this estimate was based on 
somewhat wrong value of the critical decay exponent).

When $\varepsilon<0$ and $\sigma=0$, the BAWLA does not belong to the DI class~\cite{DR2019,Park2020B}.
For example, the critical decay exponent for this case is $\delta_0 = 0.2394$~\cite{Park2020B},
which differs clearly from $\delta_\text{DI} = 0.2872$. 
Let us first check whether the BAWLA with $\sigma=0$ indeed forms a universality class.
In Fig.~\ref{Fig:di}(b), we present simulation results 
for $\varepsilon = -0.16$ at $q=0.612~25$, 0.612~27, and 0.612~29.
The system size in the simulation is $L=2^{21}$ and the numbers of independent runs 
are 160, 320, 160 for $q=0.612~25$, 0.612~27, 0.612~29, respectively.
We depict $\rho t^{\delta_0}$ as a function of $t$. At $q=0.612~27$,
the curve is almost flat for more than two logarithmic decades,
which gives $q_c = 0.612~27(2)$ and confirms that the critical behavior of the
BAWLA with $\sigma=0$ is indeed universal.

\begin{figure}
	\includegraphics[width=\linewidth]{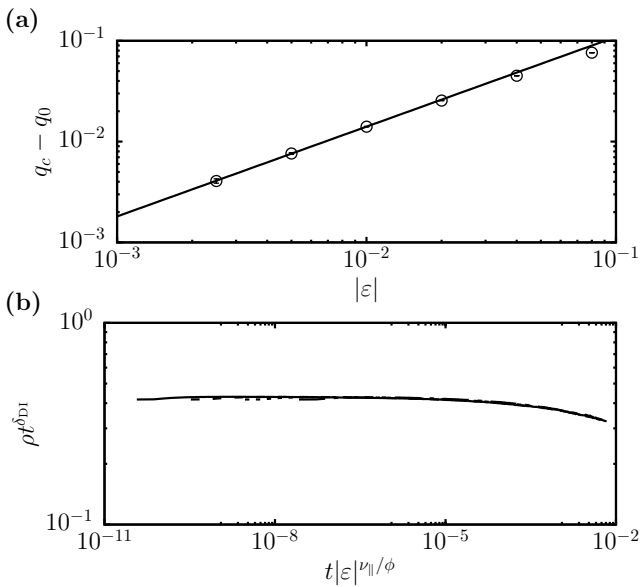}
	\caption{\label{Fig:pb0} (a) Plot of $q_c - q_0$ vs. $|\varepsilon|$
	for the BAWLA with $\sigma=0$ and $\ell=2$ on a double logarithmic
	scale. The size of error bars are smaller than the symbol size.
	The straight line with slope 0.89 shows the result of
	power-law fitting.
	(b) Scaling collapse plot of $\rho t^{\delta_\text{DI}}$ vs.
	$t |\varepsilon|^{\nu_\|/\phi}$ at $q = q_0=0.489~531$ for
	$\varepsilon = -5\times 10^{-4}$, $-10^{-3}$, $-2\times 10^{-3}$,
	and $-4\times 10^{-3}$ on a double logarithmic scale.
	All data collapse nicely onto a single curve for large $t$.
	}
\end{figure}

We found critical points by extensive Monte Carlo simulations for 
various $\varepsilon$'s, which are tabulated in Table~\ref{Tab:pc}.
In Fig.~\ref{Fig:pb0}(a), we plot $q_c - q_0$ as a function of 
$|\varepsilon|$ on a double logarithmic scale. 
We fit the data for small $|\varepsilon|$ using a power-law function
to get $1/\phi = 0.89(1)$, or $\phi = 1.123(13)$.

If $q$ is fixed to be $q_0$, the scaling function~\eqref{Eq:cross} suggests that
plots of $\rho t^{\delta_\text{DI}}$ against $t |\varepsilon|^{\nu_\|/\phi}$
should collapse onto a single curve for small $|\varepsilon|$.
Using $\nu_\| = 3.55$~\cite{Park2020}, we get $\nu_\|/\phi \approx 3.15$.
A scaling-collapse plot can be used to check the consistency of the estimate of $\phi$.
Figure~\ref{Fig:pb0}(b) shows such a scaling-collapse plot for
$\varepsilon = -5\times 10^{-4}$, $-10^{-3}$, $-2\times 10^{-3}$,
and $-4\times 10^{-3}$. The system size in the simulation is $L=2^{23}$
and the number of independent runs for the smallest $|\varepsilon|$ is 400.
Indeed, the data are well collapsed onto a single curve, which also indirectly supports
the recent estimate of $\nu_\|$ in Ref.~\cite{Park2020}.

It is worthwhile to comment on another form of attraction that can be found in the literature.
The bias due to long-range attraction with $\sigma=0$ superficially looks similar to a bias
due to a symmetry breaking field introduced to a DI model~\cite{KHP1999}.
To be concrete, let us consider a mapping from the BAW with $\ell=2$ to a nonequilibrium kinetic Ising model (NEKIM) such that a particle in the BAW is interpreted as a domain wall in the kinetic Ising model.
Details of the mapping are summarized in Table~\ref{Tab:nekim}.
We only consider the initial condition with ordering $\ldots A_+A_-A_+A_-\ldots$,
which is preserved by dynamics.
Note that the BAW in this paper corresponds to the rate $D_1=D_2 = p/2$
and $\lambda = (1-p)/2$.

In the NEKIM, there are two absorbing states with all spins either up or down.
If $D_1 > D_2$, then hopping of $A_+$ ($A_-$) particle are more likely to hop 
to the right (left), which effectively enlarges a domain of upspins in the NEKIM.
This bias can be interpreted as an effect of external magnetic field coupled to Ising spins,
which breaks the symmetry between the two absorbing states.
In the presence of the symmetry breaking field, the model now belongs to the DP class
and its crossover exponent is $1/\phi\approx 0.47$ or $\phi \approx 2.1$~\cite{BB1996,KHP1999,OM2008,PP2008PRE}.

Since the symmetry between two absorbing states (when interpreted in terms of a equivalent NEKIM) 
is not broken by long-range attraction, the bias due to a symmetry breaking field should be regarded
as a different perturbation from that due to long-range interaction.
This can be clearly explained by the different values of  the corresponding crossover exponents,
$1/\phi=0.89$ for the long-range attraction  and $1/\phi = 0.47$ for the symmetry breaking field.
Besides, changing the direction of the external magnetic field ($h \mapsto -h$) does not induce
new phenomena, because the dynamics are invariant under the simultaneous transformations
$h \mapsto -h$ and $A_+ \leftrightarrow A_-$. This should be contrasted with sign change of $\varepsilon$ in the BAWL, which has a drastic effect as we observed.

\begin{table}
	\caption{\label{Tab:nekim} A mapping of the BAW ($\ell=2$) to a nonequilibrium kinetic Ising model.  $\uparrow$ and $\downarrow$ stand for Ising spins with spin 1 and $-1$, respectively.
	$A_+$ and $A_-$ are interpreted as domain walls $\uparrow \downarrow$ and $\downarrow \uparrow$,
	respectively. 
	In branching events, a parent is indicated by a bold face symbol.
	If $D_1=D_2$, $A_+$ and $A_-$ are dynamically indistinguishable.
	}
\begin{ruledtabular}
	\begin{tabular}{ccc}
		BAW&NEKIM&transition rate\\
		\hline
		$A_+ \emptyset \Rightarrow \emptyset A_+  $ & $\uparrow \downarrow \downarrow \Rightarrow \uparrow\uparrow \downarrow$&$D_1$\\
		$\emptyset A_- \Rightarrow A_- \emptyset  $ & $\downarrow \downarrow \uparrow \Rightarrow \downarrow\uparrow \uparrow$&$D_1$\\
		$A_- \emptyset \Rightarrow \emptyset A_-  $ & $\downarrow \uparrow \uparrow \Rightarrow \downarrow\downarrow \uparrow$&$D_2$\\
		$\emptyset A_+ \Rightarrow A_+ \emptyset  $ & $\uparrow \uparrow \downarrow \Rightarrow \uparrow\downarrow \downarrow$&$D_2$\\
		$A_+ A_- \Rightarrow \emptyset \emptyset  $ & $\uparrow \downarrow \uparrow \Rightarrow \uparrow\uparrow \uparrow$&$2D_1$\\
		$A_- A_+ \Rightarrow \emptyset \emptyset  $ & $\downarrow \uparrow \downarrow \Rightarrow \downarrow\downarrow \downarrow$&$2D_2$\\
		$\bm{A_+} \emptyset  \emptyset  \Leftrightarrow \bm{A_+}A_-A_+$ & $\uparrow \downarrow \downarrow \downarrow \Leftrightarrow \uparrow \downarrow \uparrow \downarrow$&$\lambda$\\
		$\bm{A_-} \emptyset  \emptyset  \Leftrightarrow \bm{A_-}A_+A_-$ & $\downarrow \uparrow \uparrow \uparrow \Leftrightarrow \downarrow \uparrow \downarrow \uparrow$&$\lambda$\\
		$\bm{A_+} A_- \emptyset  \Leftrightarrow \bm{A_+}\emptyset  A_-$ & $\uparrow \downarrow \uparrow \uparrow \Leftrightarrow \uparrow \downarrow \downarrow\uparrow $&$\lambda$\\
		$\bm{A_-} A_+ \emptyset  \Leftrightarrow \bm{A_-}\emptyset  A_+$ & $\downarrow \uparrow \downarrow \downarrow \Leftrightarrow \downarrow \uparrow \uparrow\downarrow $&$\lambda$\\
		$\emptyset  \emptyset  \bm{A_+} \Leftrightarrow {A_+}A_-\bm{A_+}$ & $\uparrow \uparrow \uparrow \downarrow \Leftrightarrow \uparrow \downarrow \uparrow \downarrow$&$\lambda$\\
		$\emptyset  \emptyset \bm{A_-}  \Leftrightarrow {A_-}A_+\bm{A_-}$ & $\downarrow \downarrow \uparrow \uparrow \Leftrightarrow \downarrow \uparrow \downarrow \uparrow$&$\lambda$\\
		$ \emptyset{A_-} \bm{A_+}  \Leftrightarrow A_-\emptyset  \bm{A_+}$ & $\downarrow \downarrow \uparrow \downarrow \Leftrightarrow \downarrow \uparrow \uparrow\downarrow $&$\lambda$\\
		$ \emptyset {A_+} \bm{A_-} \Leftrightarrow {A_+}\emptyset  \bm{A_-}$ & $\uparrow \uparrow \downarrow \uparrow \Leftrightarrow \uparrow \downarrow \downarrow\uparrow $&$\lambda$\\
	\end{tabular}
\end{ruledtabular}
\end{table}
\subsection{\label{Sec:nbawr2}Reentrant phase transition for $\sigma=1$}
\begin{figure}
	\includegraphics[width=\linewidth]{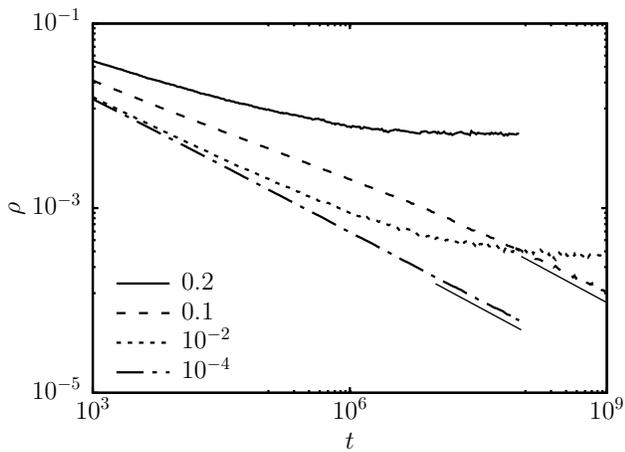}
	\caption{\label{Fig:evenre} Double-logarithmic plots of $\rho$ vs. $t$
	for the BAWLR with $\sigma=1$, $\ell=2$, $\mu=0.2$,
	$\varepsilon=0.35$ for $q=10^{-4}$ (dot-dashed line),
	$10^{-2}$ (dotted line), $0.1$ (dashed line), and 0.2 (solid line).
	Line segments with slope 0.5 are for guides to the eyes.
	}
\end{figure}

Now we move on to the BAWLR with $\sigma=1$ and $\varepsilon<0.5$.  
As shown in Fig.~\ref{Fig:dis1}, this case belongs to the DI class.
An intriguing phenomenon was observed as $q$ varies
for $\varepsilon = 0.35$ and $\mu=0.2$.
In Fig.~\ref{Fig:evenre}, we depict the behavior of the density
for $q=10^{-4}$, $10^{-2}$, 0.1, and 0.2.
Since we have found $q_c \approx 0.165$ in Fig.~\ref{Fig:dis1},
it is natural to expect that the density decays as $t^{-0.5}$
for $q=0.1$ and saturates for $q=0.2$, which is indeed the case.
A rather surprising result is the behavior for $q=10^{-2}$,
which has nonzero steady-state density. 
Since the absorbing state is stable for very small $q$ as the curve for $q=10^{-4}$ in Fig.~\ref{Fig:evenre}
shows, there should be different transitions points in three regions, 
$0<q<0.01$, $0.01<q<0.1$, and $0.1<q<0.2$.
We did not dare to find the phase boundary with accuracy. Our preliminary simulations 
showed that the reentrant phase transitions occur for some narrow region of $\varepsilon$, which is
roughly $0.34<\varepsilon<0.36$.

We think the reentrant phase transitions should be accounted for by the nonattractiveness of the BAWLR as
in Sec.~\ref{Sec:odd}. For extremely small $q$, the discussion in Sec.~\ref{Sec:smallp} shows that
the absorbing state is stable. As $q$ increases, the steady-state density starts to be nonzero
from a certain $q_1$. And then competition between the nonattractiveness and the branching 
results in a reentrant phase transitions in the region $q>q_1$.
But we do not have a good quantitative argument about the region of $\varepsilon$, where
the reentrant phase transitions occur. We would like to leave this intriguing question to later research. 

\section{\label{Sec:sum}Summary}
In this paper, we have studied absorbing phase transitions in the branching annihilating random walk
with long-range interaction with main focus on the effect of repulsion.
The results in Ref.~\cite{Pun}, especially $P_s$, are importantly relied on to understand the behavior of the BAWL.

When the number $\ell$ of offspring per branching is $1$, 
the condition that the absorbing state is stable against small perturbation of
branching is found to be $P_s\le \frac{1}{2}$.
Even though $P_s > \frac{1}{2}$, there can be reentrant phase transitions
from the active phase, to the absorbing phase, and back to the active phase
as the branching rate $q$ increases from zero. 
This reentrant phase transitions are explained, based on the nonattractiveness~\cite{Bramson1985} of the BAWLR.
The critical phenomena at nontrivial transition point are described by the DP scaling.
We also argued that a similar conclusion should be arrived at for any odd number of $\ell$.

When $\ell=2$, unlike the case with $\ell = 1$, there is only the active phase for $q>0$ 
if $P_s > 0$. Only when $P_s = 0$, the transition point $q_c$ can be nonzero.
For the case with $P_s = 0$, we studied two cases; $\sigma=0$ with $\varepsilon<0$ and $\sigma=1$
with $\varepsilon<\frac{1}{2}$.
For the first case, we studied the crossover from the DI class to the class with $\sigma=0$.
We found the crossover exponent to be $\phi =1.123(13)$.
For the second case, we have observed reentrant phase transitions that occur in the order; absorbing, active, absorbing, and then active phases. All the three transitions share critical behavior with the DI class.

When $q_c = 0$, we suggested  the scaling functions around $q=0$.
The scaling function for $\sigma<1$ is different from that for $\sigma=1$ in that
a logarithmic scaling appears for $\sigma<1$; see Eqs.~\eqref{Eq:scale} and \eqref{Eq:s1scale}
for $\sigma<1$ and $\sigma=1$, respectively.
These scaling functions are confirmed by numerical simulations.
\begin{acknowledgments}
This work was supported by the National Research Foundation of Korea (NRF) grant funded by the Korea government (MSIT) (No. 2020R1F1A1077065). 
\end{acknowledgments}
\bibliography{Park}

\begin{thebibliography}{33}%
\makeatletter
\providecommand \@ifxundefined [1]{%
 \@ifx{#1\undefined}
}%
\providecommand \@ifnum [1]{%
 \ifnum #1\expandafter \@firstoftwo
 \else \expandafter \@secondoftwo
 \fi
}%
\providecommand \@ifx [1]{%
 \ifx #1\expandafter \@firstoftwo
 \else \expandafter \@secondoftwo
 \fi
}%
\providecommand \natexlab [1]{#1}%
\providecommand \enquote  [1]{``#1''}%
\providecommand \bibnamefont  [1]{#1}%
\providecommand \bibfnamefont [1]{#1}%
\providecommand \citenamefont [1]{#1}%
\providecommand \href@noop [0]{\@secondoftwo}%
\providecommand \href [0]{\begingroup \@sanitize@url \@href}%
\providecommand \@href[1]{\@@startlink{#1}\@@href}%
\providecommand \@@href[1]{\endgroup#1\@@endlink}%
\providecommand \@sanitize@url [0]{\catcode `\\12\catcode `\$12\catcode
  `\&12\catcode `\#12\catcode `\^12\catcode `\_12\catcode `\%12\relax}%
\providecommand \@@startlink[1]{}%
\providecommand \@@endlink[0]{}%
\providecommand \url  [0]{\begingroup\@sanitize@url \@url }%
\providecommand \@url [1]{\endgroup\@href {#1}{\urlprefix }}%
\providecommand \urlprefix  [0]{URL }%
\providecommand \Eprint [0]{\href }%
\providecommand \doibase [0]{https://doi.org/}%
\providecommand \selectlanguage [0]{\@gobble}%
\providecommand \bibinfo  [0]{\@secondoftwo}%
\providecommand \bibfield  [0]{\@secondoftwo}%
\providecommand \translation [1]{[#1]}%
\providecommand \BibitemOpen [0]{}%
\providecommand \bibitemStop [0]{}%
\providecommand \bibitemNoStop [0]{.\EOS\space}%
\providecommand \EOS [0]{\spacefactor3000\relax}%
\providecommand \BibitemShut  [1]{\csname bibitem#1\endcsname}%
\let\auto@bib@innerbib\@empty
\bibitem [{\citenamefont {Bramson}\ and\ \citenamefont
  {Gray}(1985)}]{Bramson1985}%
  \BibitemOpen
  \bibfield  {author} {\bibinfo {author} {\bibfnamefont {M.}~\bibnamefont
  {Bramson}}\ and\ \bibinfo {author} {\bibfnamefont {L.}~\bibnamefont {Gray}},\
  }\bibfield  {title} {\bibinfo {title} {{The survival of branching
  annihilating random walk}},\ }\href@noop {} {\bibfield  {journal} {\bibinfo
  {journal} {Z. Wahrsch. verw. Gebiete}\ }\textbf {\bibinfo {volume} {68}},\
  \bibinfo {pages} {447} (\bibinfo {year} {1985})}\BibitemShut {NoStop}%
\bibitem [{\citenamefont {Sudbury}(1990)}]{Sudbury1990}%
  \BibitemOpen
  \bibfield  {author} {\bibinfo {author} {\bibfnamefont {A.}~\bibnamefont
  {Sudbury}},\ }\bibfield  {title} {\bibinfo {title} {{The branching
  annihilating process: an interacting particle system}},\ }\href@noop {}
  {\bibfield  {journal} {\bibinfo  {journal} {Ann. Probab.}\ }\textbf {\bibinfo
  {volume} {18}},\ \bibinfo {pages} {581} (\bibinfo {year} {1990})}\BibitemShut
  {NoStop}%
\bibitem [{\citenamefont {Takayasu}\ and\ \citenamefont
  {Tretyakov}(1992)}]{TT1992}%
  \BibitemOpen
  \bibfield  {author} {\bibinfo {author} {\bibfnamefont {H.}~\bibnamefont
  {Takayasu}}\ and\ \bibinfo {author} {\bibfnamefont {A.~Y.}\ \bibnamefont
  {Tretyakov}},\ }\bibfield  {title} {\bibinfo {title} {Extinction, survival,
  and dynamical phase transition of branching annihilating random walk},\
  }\href@noop {} {\bibfield  {journal} {\bibinfo  {journal} {Phys. Rev. Lett.}\
  }\textbf {\bibinfo {volume} {68}},\ \bibinfo {pages} {3060} (\bibinfo {year}
  {1992})}\BibitemShut {NoStop}%
\bibitem [{\citenamefont {Jensen}(1993{\natexlab{a}})}]{J1993bawo}%
  \BibitemOpen
  \bibfield  {author} {\bibinfo {author} {\bibfnamefont {I.}~\bibnamefont
  {Jensen}},\ }\bibfield  {title} {\bibinfo {title} {Critical behavior of
  branching annihilating random walks with an odd number of offsprings},\
  }\href {https://doi.org/10.1103/PhysRevE.47.R1} {\bibfield  {journal}
  {\bibinfo  {journal} {Phys. Rev. E}\ }\textbf {\bibinfo {volume} {47}},\
  \bibinfo {pages} {R1} (\bibinfo {year} {1993}{\natexlab{a}})}\BibitemShut
  {NoStop}%
\bibitem [{\citenamefont {Jensen}(1993{\natexlab{b}})}]{Jensen1993JPA}%
  \BibitemOpen
  \bibfield  {author} {\bibinfo {author} {\bibfnamefont {I.}~\bibnamefont
  {Jensen}},\ }\bibfield  {title} {\bibinfo {title} {Conservation laws and
  universality in branching annihilating random walks},\ }\href@noop {}
  {\bibfield  {journal} {\bibinfo  {journal} {J. Phys. A: Math. Gen.}\ }\textbf
  {\bibinfo {volume} {26}},\ \bibinfo {pages} {3921} (\bibinfo {year}
  {1993}{\natexlab{b}})}\BibitemShut {NoStop}%
\bibitem [{\citenamefont {Jensen}(1994)}]{J1994}%
  \BibitemOpen
  \bibfield  {author} {\bibinfo {author} {\bibfnamefont {I.}~\bibnamefont
  {Jensen}},\ }\bibfield  {title} {\bibinfo {title} {Critical exponents for
  branching annihilating random walks with an even number of offspring},\
  }\href@noop {} {\bibfield  {journal} {\bibinfo  {journal} {Phys. Rev. E}\
  }\textbf {\bibinfo {volume} {50}},\ \bibinfo {pages} {3623} (\bibinfo {year}
  {1994})}\BibitemShut {NoStop}%
\bibitem [{\citenamefont {Janssen}(1981)}]{J1981}%
  \BibitemOpen
  \bibfield  {author} {\bibinfo {author} {\bibfnamefont {H.-K.}\ \bibnamefont
  {Janssen}},\ }\bibfield  {title} {\bibinfo {title} {On the nonequilibrium
  phase transition in reaction-diffusion systems with an absorbing stationary
  state},\ }\href {https://doi.org/10.1007/BF01319549} {\bibfield  {journal}
  {\bibinfo  {journal} {Z. Phys. B}\ }\textbf {\bibinfo {volume} {42}},\
  \bibinfo {pages} {151} (\bibinfo {year} {1981})}\BibitemShut {NoStop}%
\bibitem [{\citenamefont {Grassberger}(1982)}]{G1982}%
  \BibitemOpen
  \bibfield  {author} {\bibinfo {author} {\bibfnamefont {P.}~\bibnamefont
  {Grassberger}},\ }\bibfield  {title} {\bibinfo {title} {On phase transitions
  in \uppercase{S}chl\"ogl's second model},\ }\href@noop {} {\bibfield
  {journal} {\bibinfo  {journal} {Z. Phys. B}\ }\textbf {\bibinfo {volume}
  {47}},\ \bibinfo {pages} {365} (\bibinfo {year} {1982})}\BibitemShut
  {NoStop}%
\bibitem [{\citenamefont {Menyh{\'a}rd}(1994)}]{M1994}%
  \BibitemOpen
  \bibfield  {author} {\bibinfo {author} {\bibfnamefont {N.}~\bibnamefont
  {Menyh{\'a}rd}},\ }\bibfield  {title} {\bibinfo {title} {One-dimensional
  non-equilibrium kinetic \uppercase{I}sing models with branching annihilating
  random walk},\ }\href@noop {} {\bibfield  {journal} {\bibinfo  {journal} {J.
  Phys. A}\ }\textbf {\bibinfo {volume} {27}},\ \bibinfo {pages} {6139}
  (\bibinfo {year} {1994})}\BibitemShut {NoStop}%
\bibitem [{\citenamefont {Hwang}\ \emph {et~al.}(1998)\citenamefont {Hwang},
  \citenamefont {Kwon}, \citenamefont {Park},\ and\ \citenamefont
  {Park}}]{HKPP1998}%
  \BibitemOpen
  \bibfield  {author} {\bibinfo {author} {\bibfnamefont {W.~M.}\ \bibnamefont
  {Hwang}}, \bibinfo {author} {\bibfnamefont {S.}~\bibnamefont {Kwon}},
  \bibinfo {author} {\bibfnamefont {H.}~\bibnamefont {Park}},\ and\ \bibinfo
  {author} {\bibfnamefont {H.}~\bibnamefont {Park}},\ }\bibfield  {title}
  {\bibinfo {title} {Critical phenomena of nonequilibrium dynamical systems
  with two absorbing states},\ }\href
  {https://doi.org/10.1103/PhysRevE.57.6438} {\bibfield  {journal} {\bibinfo
  {journal} {Phys. Rev. E}\ }\textbf {\bibinfo {volume} {57}},\ \bibinfo
  {pages} {6438} (\bibinfo {year} {1998})}\BibitemShut {NoStop}%
\bibitem [{\citenamefont {Hinrichsen}(2000)}]{H2000}%
  \BibitemOpen
  \bibfield  {author} {\bibinfo {author} {\bibfnamefont {H.}~\bibnamefont
  {Hinrichsen}},\ }\bibfield  {title} {\bibinfo {title} {Non-equilibrium
  critical phenomena and phase transitions into absorbing states},\ }\href@noop
  {} {\bibfield  {journal} {\bibinfo  {journal} {Adv. Phys.}\ }\textbf
  {\bibinfo {volume} {49}},\ \bibinfo {pages} {815} (\bibinfo {year}
  {2000})}\BibitemShut {NoStop}%
\bibitem [{\citenamefont {\'{O}dor}(2004)}]{O2004}%
  \BibitemOpen
  \bibfield  {author} {\bibinfo {author} {\bibfnamefont {G.}~\bibnamefont
  {\'{O}dor}},\ }\bibfield  {title} {\bibinfo {title} {Universality classes in
  nonequilibrium lattice systems},\ }\href
  {https://doi.org/10.1103/RevModPhys.76.663} {\bibfield  {journal} {\bibinfo
  {journal} {Rev. Mod. Phys.}\ }\textbf {\bibinfo {volume} {76}},\ \bibinfo
  {eid} {663} (\bibinfo {year} {2004})}\BibitemShut {NoStop}%
\bibitem [{\citenamefont {Henkel}\ \emph {et~al.}(2008)\citenamefont {Henkel},
  \citenamefont {Hinrichsen},\ and\ \citenamefont {L{\"u}beck}}]{HHL2008Book}%
  \BibitemOpen
  \bibfield  {author} {\bibinfo {author} {\bibfnamefont {M.}~\bibnamefont
  {Henkel}}, \bibinfo {author} {\bibfnamefont {H.}~\bibnamefont {Hinrichsen}},\
  and\ \bibinfo {author} {\bibfnamefont {S.}~\bibnamefont {L{\"u}beck}},\
  }\href@noop {} {\emph {\bibinfo {title} {Non-Equilibrium Phase Transitions:
  Volume 1: Absorbing Phase Transitions}}}\ (\bibinfo  {publisher} {Springer},\
  \bibinfo {address} {Berlin},\ \bibinfo {year} {2008})\BibitemShut {NoStop}%
\bibitem [{\citenamefont {Grassberger}\ \emph {et~al.}(1984)\citenamefont
  {Grassberger}, \citenamefont {Krause},\ and\ \citenamefont {von~der
  Twer}}]{GKvdT1984}%
  \BibitemOpen
  \bibfield  {author} {\bibinfo {author} {\bibfnamefont {P.}~\bibnamefont
  {Grassberger}}, \bibinfo {author} {\bibfnamefont {F.}~\bibnamefont
  {Krause}},\ and\ \bibinfo {author} {\bibfnamefont {T.}~\bibnamefont {von~der
  Twer}},\ }\bibfield  {title} {\bibinfo {title} {A new type of kinetic
  critical phenomenon},\ }\href@noop {} {\bibfield  {journal} {\bibinfo
  {journal} {J. Phys. A}\ }\textbf {\bibinfo {volume} {17}},\ \bibinfo {pages}
  {L105} (\bibinfo {year} {1984})}\BibitemShut {NoStop}%
\bibitem [{\citenamefont {Jensen}(1993{\natexlab{c}})}]{J1993}%
  \BibitemOpen
  \bibfield  {author} {\bibinfo {author} {\bibfnamefont {I.}~\bibnamefont
  {Jensen}},\ }\bibfield  {title} {\bibinfo {title} {Critical behavior of the
  pair contact process},\ }\href@noop {} {\bibfield  {journal} {\bibinfo
  {journal} {Phys. Rev. Lett.}\ }\textbf {\bibinfo {volume} {70}},\ \bibinfo
  {pages} {1465} (\bibinfo {year} {1993}{\natexlab{c}})}\BibitemShut {NoStop}%
\bibitem [{\citenamefont {Cardy}\ and\ \citenamefont
  {T{\"a}uber}(1998)}]{CT1998}%
  \BibitemOpen
  \bibfield  {author} {\bibinfo {author} {\bibfnamefont {J.~L.}\ \bibnamefont
  {Cardy}}\ and\ \bibinfo {author} {\bibfnamefont {U.~C.}\ \bibnamefont
  {T{\"a}uber}},\ }\bibfield  {title} {\bibinfo {title} {Field theory of
  branching and annihilating random walks},\ }\href@noop {} {\bibfield
  {journal} {\bibinfo  {journal} {J. Stat. Phys.}\ }\textbf {\bibinfo {volume}
  {90}},\ \bibinfo {pages} {1} (\bibinfo {year} {1998})}\BibitemShut {NoStop}%
\bibitem [{\citenamefont {Al~Hammal}\ \emph {et~al.}(2005)\citenamefont
  {Al~Hammal}, \citenamefont {Chat\'{e}}, \citenamefont {Dornic},\ and\
  \citenamefont {Mu{\~n}oz}}]{HCDM2005}%
  \BibitemOpen
  \bibfield  {author} {\bibinfo {author} {\bibfnamefont {O.}~\bibnamefont
  {Al~Hammal}}, \bibinfo {author} {\bibfnamefont {H.}~\bibnamefont
  {Chat\'{e}}}, \bibinfo {author} {\bibfnamefont {I.}~\bibnamefont {Dornic}},\
  and\ \bibinfo {author} {\bibfnamefont {M.~A.}\ \bibnamefont {Mu{\~n}oz}},\
  }\bibfield  {title} {\bibinfo {title} {Langevin description of critical
  phenomena with two symmetric absorbing states},\ }\href
  {https://doi.org/10.1103/PhysRevLett.94.230601} {\bibfield  {journal}
  {\bibinfo  {journal} {Phys. Rev. Lett.}\ }\textbf {\bibinfo {volume} {94}},\
  \bibinfo {eid} {230601} (\bibinfo {year} {2005})}\BibitemShut {NoStop}%
\bibitem [{\citenamefont {Daga}\ and\ \citenamefont {Ray}(2019)}]{DR2019}%
  \BibitemOpen
  \bibfield  {author} {\bibinfo {author} {\bibfnamefont {B.}~\bibnamefont
  {Daga}}\ and\ \bibinfo {author} {\bibfnamefont {P.}~\bibnamefont {Ray}},\
  }\bibfield  {title} {\bibinfo {title} {Universality classes of absorbing
  phase transitions in generic branching-annihilating particle systems with
  nearest-neighbor bias},\ }\href {https://doi.org/10.1103/PhysRevE.99.032104}
  {\bibfield  {journal} {\bibinfo  {journal} {Phys. Rev. E}\ }\textbf {\bibinfo
  {volume} {99}},\ \bibinfo {pages} {032104} (\bibinfo {year}
  {2019})}\BibitemShut {NoStop}%
\bibitem [{\citenamefont {Park}(2020{\natexlab{a}})}]{Park2020B}%
  \BibitemOpen
  \bibfield  {author} {\bibinfo {author} {\bibfnamefont {S.-C.}\ \bibnamefont
  {Park}},\ }\bibfield  {title} {\bibinfo {title} {Branching annihilating
  random walks with long-range attraction in one dimension},\ }\href@noop {}
  {\bibfield  {journal} {\bibinfo  {journal} {Phys. Rev. E}\ }\textbf {\bibinfo
  {volume} {101}},\ \bibinfo {pages} {052125} (\bibinfo {year}
  {2020}{\natexlab{a}})}\BibitemShut {NoStop}%
\bibitem [{\citenamefont {Park}(2020{\natexlab{b}})}]{Park2020A}%
  \BibitemOpen
  \bibfield  {author} {\bibinfo {author} {\bibfnamefont {S.-C.}\ \bibnamefont
  {Park}},\ }\bibfield  {title} {\bibinfo {title} {Crossover behaviors in
  branching annihilating attracting walk},\ }\href@noop {} {\bibfield
  {journal} {\bibinfo  {journal} {Phys. Rev. E}\ }\textbf {\bibinfo {volume}
  {101}},\ \bibinfo {pages} {052103} (\bibinfo {year}
  {2020}{\natexlab{b}})}\BibitemShut {NoStop}%
\bibitem [{\citenamefont {Park}(ev E)}]{Pun}%
  \BibitemOpen
  \bibfield  {author} {\bibinfo {author} {\bibfnamefont {S.-C.}\ \bibnamefont
  {Park}},\ }\href@noop {} {\bibinfo {title} {One-dimensional annihilating
  random walk with long-range interaction}} (\bibinfo {year} {to appear in
  Phys. Rev. E}),\ \Eprint {https://arxiv.org/abs/arXiv:2007.08748}
  {arXiv:2007.08748} \BibitemShut {NoStop}%
\bibitem [{\citenamefont {Sen}\ and\ \citenamefont {Ray}(2015)}]{Sen2015}%
  \BibitemOpen
  \bibfield  {author} {\bibinfo {author} {\bibfnamefont {P.}~\bibnamefont
  {Sen}}\ and\ \bibinfo {author} {\bibfnamefont {P.}~\bibnamefont {Ray}},\
  }\bibfield  {title} {\bibinfo {title} {\uppercase{$A+A \rightarrow
  \emptyset$} model with a bias towards nearest neighbor},\ }\href@noop {}
  {\bibfield  {journal} {\bibinfo  {journal} {Phys. Rev. E}\ }\textbf {\bibinfo
  {volume} {92}},\ \bibinfo {pages} {012109} (\bibinfo {year}
  {2015})}\BibitemShut {NoStop}%
\bibitem [{\citenamefont {Roy}\ and\ \citenamefont {Sen}(shed)}]{Roy2020}%
  \BibitemOpen
  \bibfield  {author} {\bibinfo {author} {\bibfnamefont {R.}~\bibnamefont
  {Roy}}\ and\ \bibinfo {author} {\bibfnamefont {P.}~\bibnamefont {Sen}},\
  }\href@noop {} {\bibinfo {title} {{$A+A\rightarrow \emptyset$ reaction for
  particles with a dynamic bias to move away from their nearest neighbour in
  one dimension}}} (\bibinfo {year} {unpublished}),\ \Eprint
  {https://arxiv.org/abs/arXiv:2004.01472} {arXiv:2004.01472} \BibitemShut
  {NoStop}%
\bibitem [{\citenamefont {Jensen}(1999)}]{J1999}%
  \BibitemOpen
  \bibfield  {author} {\bibinfo {author} {\bibfnamefont {I.}~\bibnamefont
  {Jensen}},\ }\bibfield  {title} {\bibinfo {title} {Low-density series
  expansions for directed percolation: I. a new efficient algorithm with
  applications to the square lattice},\ }\href@noop {} {\bibfield  {journal}
  {\bibinfo  {journal} {J. Phys. A}\ }\textbf {\bibinfo {volume} {32}},\
  \bibinfo {pages} {5233} (\bibinfo {year} {1999})}\BibitemShut {NoStop}%
\bibitem [{\citenamefont {Park}(2013)}]{Park2013}%
  \BibitemOpen
  \bibfield  {author} {\bibinfo {author} {\bibfnamefont {S.-C.}\ \bibnamefont
  {Park}},\ }\bibfield  {title} {\bibinfo {title} {High-precision estimate of
  the critical exponents for the directed \uppercase{I}sing universality
  class},\ }\href@noop {} {\bibfield  {journal} {\bibinfo  {journal} {J. Korean
  Phys. Soc.}\ }\textbf {\bibinfo {volume} {62}},\ \bibinfo {pages} {469}
  (\bibinfo {year} {2013})}\BibitemShut {NoStop}%
\bibitem [{\citenamefont {Harris}(1974)}]{H1974}%
  \BibitemOpen
  \bibfield  {author} {\bibinfo {author} {\bibfnamefont {T.~E.}\ \bibnamefont
  {Harris}},\ }\bibfield  {title} {\bibinfo {title} {Contact interactions on a
  lattice},\ }\href@noop {} {\bibfield  {journal} {\bibinfo  {journal} {Ann.
  Prob.}\ }\textbf {\bibinfo {volume} {2}},\ \bibinfo {pages} {969} (\bibinfo
  {year} {1974})}\BibitemShut {NoStop}%
\bibitem [{\citenamefont {Liggett}(1985)}]{Liggett1985Book}%
  \BibitemOpen
  \bibfield  {author} {\bibinfo {author} {\bibfnamefont {T.~M.}\ \bibnamefont
  {Liggett}},\ }\href@noop {} {\emph {\bibinfo {title} {Interacting particle
  systems}}}\ (\bibinfo  {publisher} {Springer},\ \bibinfo {address} {Berlin},\
  \bibinfo {year} {1985})\BibitemShut {NoStop}%
\bibitem [{\citenamefont {Kwon}\ and\ \citenamefont {Park}(1995)}]{KP1995}%
  \BibitemOpen
  \bibfield  {author} {\bibinfo {author} {\bibfnamefont {S.}~\bibnamefont
  {Kwon}}\ and\ \bibinfo {author} {\bibfnamefont {H.}~\bibnamefont {Park}},\
  }\bibfield  {title} {\bibinfo {title} {Reentrant phase diagram of branching
  annihilating random walks with one and two offspring},\ }\href@noop {}
  {\bibfield  {journal} {\bibinfo  {journal} {Phys. Rev. E}\ }\textbf {\bibinfo
  {volume} {52}},\ \bibinfo {pages} {5955} (\bibinfo {year}
  {1995})}\BibitemShut {NoStop}%
\bibitem [{\citenamefont {Park}(2020{\natexlab{c}})}]{Park2020}%
  \BibitemOpen
  \bibfield  {author} {\bibinfo {author} {\bibfnamefont {S.-C.}\ \bibnamefont
  {Park}},\ }\bibfield  {title} {\bibinfo {title} {Order-parameter critical
  exponent of absorbing phase transitions in one-dimensional systems with two
  symmetric absorbing states},\ }\href@noop {} {\bibfield  {journal} {\bibinfo
  {journal} {Phys. Rev. E}\ }\textbf {\bibinfo {volume} {101}},\ \bibinfo
  {pages} {052114} (\bibinfo {year} {2020}{\natexlab{c}})}\BibitemShut
  {NoStop}%
\bibitem [{\citenamefont {Park}\ and\ \citenamefont {Park}(2008)}]{PP2008PRE}%
  \BibitemOpen
  \bibfield  {author} {\bibinfo {author} {\bibfnamefont {S.-C.}\ \bibnamefont
  {Park}}\ and\ \bibinfo {author} {\bibfnamefont {H.}~\bibnamefont {Park}},\
  }\bibfield  {title} {\bibinfo {title} {Three different routes from the
  directed \uppercase{I}sing to the directed percolation class},\ }\href@noop
  {} {\bibfield  {journal} {\bibinfo  {journal} {Phys. Rev. E}\ }\textbf
  {\bibinfo {volume} {78}},\ \bibinfo {pages} {041128} (\bibinfo {year}
  {2008})}\BibitemShut {NoStop}%
\bibitem [{\citenamefont {Kwon}\ \emph {et~al.}(1999)\citenamefont {Kwon},
  \citenamefont {Hwang},\ and\ \citenamefont {Park}}]{KHP1999}%
  \BibitemOpen
  \bibfield  {author} {\bibinfo {author} {\bibfnamefont {S.}~\bibnamefont
  {Kwon}}, \bibinfo {author} {\bibfnamefont {W.~M.}\ \bibnamefont {Hwang}},\
  and\ \bibinfo {author} {\bibfnamefont {H.}~\bibnamefont {Park}},\ }\bibfield
  {title} {\bibinfo {title} {Dynamic behavior of driven interfaces in models
  with two absorbing states},\ }\href@noop {} {\bibfield  {journal} {\bibinfo
  {journal} {Phys. Rev. E}\ }\textbf {\bibinfo {volume} {59}},\ \bibinfo
  {pages} {4949} (\bibinfo {year} {1999})}\BibitemShut {NoStop}%
\bibitem [{\citenamefont {Bassler}\ and\ \citenamefont
  {Browne}(1996)}]{BB1996}%
  \BibitemOpen
  \bibfield  {author} {\bibinfo {author} {\bibfnamefont {K.~E.}\ \bibnamefont
  {Bassler}}\ and\ \bibinfo {author} {\bibfnamefont {D.~A.}\ \bibnamefont
  {Browne}},\ }\bibfield  {title} {\bibinfo {title} {Nonequilibrium critical
  dynamics of a three species monomer-monomer model},\ }\href@noop {}
  {\bibfield  {journal} {\bibinfo  {journal} {Phys. Rev. Lett.}\ }\textbf
  {\bibinfo {volume} {77}},\ \bibinfo {pages} {4094} (\bibinfo {year}
  {1996})}\BibitemShut {NoStop}%
\bibitem [{\citenamefont {\'{O}dor}\ and\ \citenamefont
  {Menyhard}(2008)}]{OM2008}%
  \BibitemOpen
  \bibfield  {author} {\bibinfo {author} {\bibfnamefont {G.}~\bibnamefont
  {\'{O}dor}}\ and\ \bibinfo {author} {\bibfnamefont {N.}~\bibnamefont
  {Menyhard}},\ }\bibfield  {title} {\bibinfo {title} {Crossovers from parity
  conserving to directed percolation universality},\ }\href@noop {} {\bibfield
  {journal} {\bibinfo  {journal} {Phys. Rev. E}\ }\textbf {\bibinfo {volume}
  {78}},\ \bibinfo {pages} {041112} (\bibinfo {year} {2008})}\BibitemShut
  {NoStop}%
\end{thebibliography}%
\end{document}